\documentclass[preprint]{aastex}
\usepackage{natbib}
\usepackage{booktabs}
\usepackage{threeparttable}

\begin{document}

\title{Observations of a quasi-periodic pulsation in the coronal loop and microwave flux during a solar preflare phase}

\author{Dong~Li\altaffilmark{1,2,3}, Ying~Li\altaffilmark{1}, Lei~Lu\altaffilmark{1}, Qingmin~Zhang\altaffilmark{1,4}, Zongjun~Ning\altaffilmark{1}, and Sergey~Anfinogentov\altaffilmark{5}}
\affil{$^1$Key Laboratory of Dark Matter and Space Astronomy, Purple Mountain Observatory, CAS, Nanjing 210033, People's Republic of China \\
     $^2$State Key Laboratory of Space Weather, Chinese Academy of Sciences, Beijing 100190, People's Republic of China \\
     $^3$CAS Key Laboratory of Solar Activity, National Astronomical Observatories, Beijing 100012, People's Republic of China \\
     $^4$State Key Laboratory of Lunar and Planetary Sciences, Macau University of Science and Technology, Macau, People's Republic of China  \\
     $^5$Institute of Solar-Terrestrial Physics (ISZF), Lermontov st., 126a, Irkutsk, 664033 - Russian Federation \\}
     \altaffiltext{}{Correspondence should be sent to: lidong@pmo.ac.cn}

\begin{abstract}
We report a quasi-periodic pulsation (QPP) event simultaneously
detected from the spatial displacements of coronal loop at both EUV
images and microwave emission during the preflare phase of a C1.1
flare on 2016 March 23. Using the motion magnification technique, a
low-amplitude transverse oscillation with the growing period is
discovered in a diffuse coronal loop in Atmospheric Imaging Assembly
(AIA) image sequences at wavelength of 171~{\AA}, and the initial
oscillation period is estimated to be $\sim$397~s with a slow growth
rate of 0.045. At the same time, a QPP with growing periods from
roughly 300~s to nearly 500~s is discovered in the microwave flux in
the same active region. Based on the imaging observations measured
at EUV wavelengths by the AIA and at microwave 17~GHz by Nobeyama
Radioheliograph, the diffuse coronal loop and the microwave
radiation source are found to be connected through a hot loop seen
in AIA images at wavelength of 94~{\AA}. The growing period of the
QPP should be related to the modulation of LRC-circuit oscillating
process in a current-carrying plasma loop. The existence of electric
currents may imply the non-potentialities in the source region
during the preflare phase.
\end{abstract}
\keywords{Solar flares --- Solar oscillations --- Solar ultraviolet
emission --- Solar radio emission}

\section{Introduction}
Quasi-periodic pulsations (QPPs) are frequently identified as
temporal oscillations of electromagnetic radiation in solar flares
\citep[see,][for reviews]{Nakariakov09,Van16}. The QPP was first
discovered by \cite{Parks69} in a solar X-ray event, and then the
observational studies of QPPs were reported in more and more
wavebands. So far, they could be observed in almost all the
wavelengths, i.e., radio/microwave, extreme-ultraviolet and
ultraviolet (EUV/UV), soft and hard X-ray (SXR/HXR), and the
detected periods can cover a broad range, which are from sub-seconds
to dozens of minutes
\citep[e.g.,][]{Tan12,Tan16,Milligan17,Kolotkov18,Nakariakov18,Hayes19,Shen19}.
Utilizing the high spatial and temporal resolution imaging and
spectroscopic observations, QPPs are also observed as spatial
motions of the oscillating system, i.e., the spatial displacement
oscillations in coronal loops
\citep{Nakariakov99,Wang12,Anfinogentov15,Yuan16,Goddard18}, the
Doppler shift oscillations in coronal or flaring loops
\citep{Ofman02,Mariska10,Tian12,Li18}. Due to the limitation of the
time and spatial resolutions of available observational facilities,
most of the spatial motions of QPPs are discovered in EUV and X-ray
images or spectral lines, and their oscillation periods are found to
be from dozens of seconds to tens of minutes
\citep{Aschwanden99,Brosius15,Tian16,Duckenfield18,Su18}. QPPs in
coronal/flaring loops often exhibit two different oscillation
behaviors, one is large amplitude but damping rapidly
\citep[e.g.,][]{Nakariakov99,Ofman02,Wang15,Li17,Nakariakov19}, and
the other one has a low amplitude without significant damping
\citep[e.g.,][]{Tian12,Nistico13,Thurgood14,Anfinogentov15,Li18}.

Based on the use of multi-instrumental and -wavelength data, the
observational study of QPPs has become an important research avenue.
To understand the generation mechanisms of the observed QPPs,
various kinds of theoretical models have been proposed
\citep[see,][for the reviews of various theoretical
models]{Nakariakov05,Nakariakov09,Van16}. In particular, the QPP
could be related to a self-oscillatory regime of spontaneous
magnetic reconnection (i.e., magnetic dripping mechanism), or it
might be driven by the magnetohydrodynamic (MHD) oscillation in
flaring plasma structures, or it is caused by the induced repetitive
magnetic reconnection which can be modulated by MHD waves, such as
fast kink waves, slow magnetoacoustic waves, and global sausage
waves
\citep[e.g.,][]{Kliem00,Chen06,Nakariakov09,Su12,Anfinogentov15,Wang15,Tian16,Goddard18}.
It is likely that different mechanisms/models generate different
types of QPPs, and it is impossible to find one physical model to
explain all the discovered QPPs \citep{Nakariakov05,Tan10,Van16}.

QPPs in coronal loops such as damping oscillations are generally
thought to be triggered by the solar eruptions, i.e., flares, jets
or EUV waves
\citep[e.g.,][]{Nakariakov99,Nakariakov05,Zimovets15,Su18}. However,
the QPP in the coronal loop during the preflare phase is rarely
reported. In this letter, using the imaging observations measured by
the Atmospheric Imaging Assembly \citep[AIA,][]{Lemen12} on board
the {\it Solar Dynamics Observatory} ({\it SDO}) and the Nobeyama
Radioheliograph \citep[NoRH,][]{Nakajima94}, we investigate a
low-amplitude transverse oscillation in the coronal loop in
AIA~171~{\AA} before a C1.1 flare. Meanwhile, a similar QPP is
detected in microwave flux at the frequency of 17~GHz. The microwave
source and the EUV coronal loop are found to be connected through a
hot loop in AIA~94~{\AA}. To the best of our knowledge, this is the
first report of QPPs simultaneously discovered from the spatial
displacements of coronal loop at both EUV images and microwave
emission during the preflare phase, which could be helpful for
understanding the precursor of a solar flare.

\section{Observation}
We analyze observations of a coronal loop in the active region of
NOAA 12524 on 2016 March 23. We focus on a low-amplitude transverse
oscillation during $\sim$02:41$-$03:01~UT in the EUV coronal loop
before a C1.1 flare which starts at about 02:59~UT in the {\it GOES}
SXR flux at 1$-$8~{\AA}. Figure~\ref{flux}~(a) shows that the light
curve in {\it GOES}~1$-$8~{\AA} (black curve) appears to increase
slowly from 02:30~UT, which is earlier than the flare onset time
(vertical black line). Moreover, the time derivative (cyan curve)
derived from {\it GOES}~1$-$8~{\AA} exhibits a clear enhancement
before the solar flare, i.e., during 02:30$-$03:01~UT. The similar
feature can be discovered in the light curves in EUV
SpectroPhotometer (ESP)~1$-$70~{\AA} (magenta curve) and its time
derivative (red cure) measured by the Extreme Ultraviolet
Variability Experiment \citep[EVE,][]{Woods12} aboard {\it SDO},
such as a slight increase in SXR flux but a prominent enhancement in
its time derivative during the same time interval. Here, the weak
enhancement observed in {\it GOES} and ESP SXR flux is regarded as
the preflare phase rather than a separate subflare \citep[see,][for
details]{Liy17}. In panel~(a), some light curves are shifted in
height to display them clearly in the same window. Finally, the
normalized microwave flux (green curve) at the frequency of 17~GHz
recorded by the NoRH also show a very weak radiation enhancement
from $\sim$02:35 to $\sim$03:01~UT. Notice that the microwave flux
is integrated over the active region such as NOAA 12524, rather than
over the full-disk Sun.

Figure~\ref{flux}~(b)$-$(d) shows the simultaneous snapshots at
around 02:50~UT with the same field-of-view (FOV) of
320$\arcsec$$\times$400$\arcsec$ in EUV and microwave channels
measured by the {\it SDO}/AIA ($\sim$0.6\arcsec/pixel) and the NoRH
($\sim$4.9\arcsec/pixel), respectively. The AIA images are
pre-processed with the standard routine of `aia\_prep.pro'
\citep{Lemen12} included in the solar software package. Panels~(b)
and (c) present the EUV images in AIA~171~{\AA} and 94~{\AA},
respectively. A number of loop structures can be seen in AIA images
at wavelength of 171~{\AA}, and a large diffuse loop with a
bubble-like profile is studied in this work, as indicated by a
yellow curve that only outlines the left part of the diffuse loop.
Two red short lines outline the slice (S1) positions used to make
the time-distance plot, while the cyan box marks the FOV of a movie
(see, 171\_94.mp4). A highly sheared arch which is close to the
diffuse coronal loop \citep[see,][]{Liy17} can be seen in AIA images
at wavelength of 171~{\AA}, as marked by a purple dashed line.
However, the diffuse coronal loop is not identified in AIA images at
wavelength of 94~{\AA}, but a faint hot loop appears above the
sheared arch (purple dashed line), as outlined by a magenta curve
(or arrow). The movie of 171\_94.mp4 further suggests that the hot
loop in AIA~94~{\AA} begins to appear at around 02:30~UT and then
becomes brighter and brighter, as marked by a magenta arrow.
Finally, an enhancement of the microwave emission at NoRH~17~GHz
(blue contours) is discovered at the end of the hot loop seen in the
AIA image at wavelength of 94~{\AA}, as shown in panels~(c) and (d).

\section{Data reduction and Result}
The movie of 171\_94.mp4 shows the temporal evolution of coronal
loops in AIA~171~{\AA} and 94~{\AA} during $\sim$02:30$-$03:00~UT
with a time cadence of 12~s. From which, we can see that a diffuse
coronal loop with bubble-like profiles rises slowly with time during
the preflare phase \citep[see details in][]{Liy17} in AIA~171~{\AA}.
The coronal loop starts to oscillate at about 02:41~UT, and could
last for nearly 20~minutes. The oscillation is transverse and
significant at the left-upper part of the coronal loop, as indicated
by two red lines. However, it is impossible to identify the
oscillation feature at the other region of the diffuse coronal loop.

Figure~\ref{slice} presents the time-distance plots along the slice
of S1, which are derived from the AIA~171~{\AA} image sequences. It
is hard to identify any apparent signature of the oscillation from
the original data but only a diffuse faint loop which is centered at
around 25\arcsec, as marked with the magenta arrows in panel~(a). To
detect the low-amplitude oscillation at the coronal loop, we then
perform a motion magnification method to the original data, which
enable us to magnify the quasi-periodic transverse motions of
contrast features with a very low amplitude such as $\sim$0.2~Mm in
image sequences, but it does not change the image scale
\citep[e.g.,][]{Anfinogentov16}. In our study, the magnification
factors of 3, 6 and 9 are applied to the AIA~171~{\AA} image
sequences, respectively. Panels~(b)$-$(d) shows time-distance plots
after the motion magnification technique. We can find a very weak
transverse oscillation in panel~(b), which displays with a
magnification factor of 3. Then, with the increasing of
magnification factors, the transverse oscillation is more  and more
clear. Finally, we perform a magnification factor of 9, and a
transverse oscillation with three pronounced peaks can be found
obviously in the time-distance plot, as indicated by the pink pluses
(`+') in panel~(d). The over-plotted green line in panel~(c)
represents the normalized microwave flux in NoRH~17~GHz measured
from the active region of NOAA~12524, as shown in
Figure~\ref{flux}~(d). The microwave flux exhibits three small-scale
peaks, which are well corresponding to the three peaks in the
transverse oscillation of coronal loop (pink pluses), suggesting
that they might come from the same heating process. However, we
could not discover the corresponding three peaks from the time
derivatives of SXR fluxes in {\it GOES}~1$-$8~{\AA} and
ESP~1-70~{\AA}, which could be due to the fact that the SXR fluxes
are measured as full-disk radiations, so the small-scale QPP
occurred in a relatively active region can not be detected.

To take a close look at the period of the low-amplitude transverse
oscillation in the EUV coronal loop, we manually extracted the
oscillation positions and applied a sine function with a linear
trend (Eq.~\ref{yfit}) to fit them, which was a widely adopted
function \citep[e.g.,][]{Nakariakov99,Zhang17,Su18}, but without the
damping term \citep[see,][]{Thurgood14,Anfinogentov15}.
\begin{equation}
  A(t) =  A_m \sin(\frac{2\pi t}{P_0+k_0 t} + \phi) + b t + A_0
 \label{yfit}
\end{equation}
Where $A_m$ is the oscillation amplitude, $A_0$ and $\phi$ represent
the initial position and initial phase, $b$ indicates the drifting
speed of oscillation loop. The oscillation period is a function of
time, $P_0$ represents the initial period, while $k_0$ refers the
changing (decreasing/growth) rate of the oscillation period.
Figure~\ref{fitp} presents the temporal evolution of the loop
locations (pluses) during the low-amplitude transverse oscillation.
Here the low-amplitude transverse oscillation has been reduced to
the real amplitude, i.e., multiple a factor of 1/9 to offset the
motion magnification technique. The best fitted curve for the loop
oscillation is shown with a red line, and the fitted results are
also given. The oscillation amplitude is about $\sim$0.8~Mm, while
the initial oscillation period is $\sim$397~s, which is slowly
growing with a growth rate of 0.045.

To examine the QPP in microwave emission, we then performed a
wavelet analysis method \citep{Torrence98} on the detrended light
curve in NoRH~17~GHz. Figure~\ref{wavlet}~(a) shows the normalized
microwave flux in NoRH~17~GHz from 02:40~UT to 03:01~UT, and the
magenta line represents the trended light curve calculated by a
420-s running average \citep[see,][]{Yuan11,Li18}. Then the
detrended light curve can be derived by subtracting the trended
light curve, as shown in panel~(b). The wavelet power spectrum shows
an obvious signature of QPP, and its period appears to increase from
$\sim$300~s to $\sim$500~s, as indicated by the red arrow in
panel~(c). The growing periods detected in microwave flux are
consistent with the growing periods in the diffuse coronal loop in
AIA~171~{\AA}. On the other hand, both the low-amplitude transverse
oscillation in the EUV coronal loop and the small-scale QPP in
microwave flux start at about 02:41~UT, which is later than the
onset time of the hot loop and the sheared arch
\citep[see,][]{Liy17}. Moreover, the hot loop and the sheared arch
connect the diffuse coronal loop and microwave source (see,
Figure~\ref{flux}~c). All these observational facts imply that the
loop oscillation and the microwave QPP could be related to each
other, and may be driven by a same process.

\section{Conclusion and Discussion}
Utilizing the observations measured by the {\it SDO}/AIA and the
NoRH, we investigated a low-amplitude transverse oscillation in the
EUV coronal loop before a {\it GOES} C1.1 flare on 2016 March 23.
The main results are summarized as follows.

\begin{enumerate}
\item A low-amplitude transverse oscillation with the growing period is discovered
in the EUV coronal loop during the preflare phase. The oscillation
period is estimated to $\sim$397~s with a growth rate of 0.045, and
the oscillation amplitude is about 0.8~Mm, without significant
damping.

\item A small-scale QPP with growing periods is detected in the microwave
flux during the same preflare phase. The quasi-periods are estimated
to increase from $\sim$300~s to $\sim$500~s.

\item The EUV diffuse coronal loop and microwave radiation source are connected
through a hot loop seen in AIA images at wavelength of 94~{\AA},
suggesting that the loop oscillation and the microwave QPP may be
triggered by a same thermal process, i.e., kink oscillations. The
growing periods might be explained by the modulation of LRC-circuit
oscillating process in a current-carrying plasma loop.
\end{enumerate}

\begin{figure}
\epsscale{1.0} \plotone{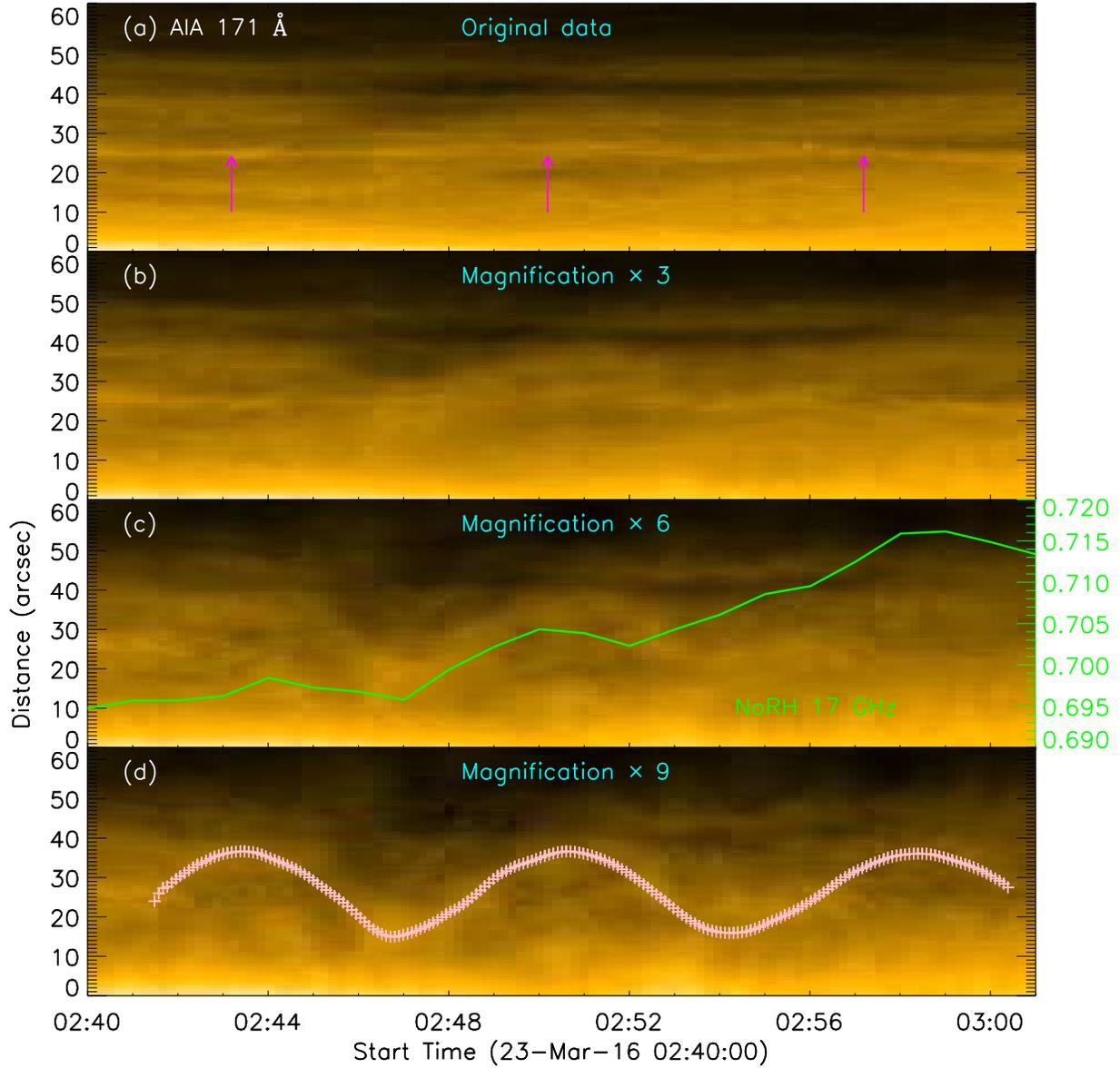} \caption{Time-distance plots along
S1 made with the original data (a) and processed with the motion
magnification technique (b$-$d). The magenta arrows mark the loop
center positions. The pink pluses (`+') mark the transverse
oscillation locations of the diffuse coronal loop. The green
over-plotted light curve represents the microwave emission
integrated from the active region in NoRH~17~GHz.} \label{slice}
\end{figure}

It is interesting that a low-amplitude transverse oscillation
without significant damping in the EUV coronal loop is observed
before the onset time of a {\it GOES} C1.1 flare. The non-damping
transverse oscillation of coronal loop can be clearly seen in AIA
image sequences at the wavelength of 171~{\AA}, see details in the
movie of 171\_94.mp4. However, it is hard to identify the signature
of oscillation in the time-distance image due to its low amplitude
such as $\sim$0.8~Mm, as shown in Figure~\ref{slice}~(a).
Fortunately, \cite{Anfinogentov16} have developed a motion
magnification technique to extract the low-amplitude transverse
oscillations in image sequences, particularly for the {\it SDO}/AIA
data cube. In this letter, we apply the motion magnification
technique to AIA~171~{\AA} image sequences, and find a pronounced
signature of low-amplitude transverse oscillation in the EUV coronal
loop (Figure~\ref{slice}~d), which agrees well with the temporal
evolution in the movie of 171\_94.mp4. The low-amplitude transverse
oscillation is non-damping, but it only lasts for three pronounced
peaks. Our observational result is different from previous
observations of non-damping loop oscillations, which are often
discovered in non-flaring active regions and can be oscillating for
dozens of cycles \citep[e.g.,][]{Tian12,Anfinogentov15,Yuan16}. The
difference might be attributed to the fact that the diffuse coronal
loop is affected seriously by a C1.1 flare after 02:59~UT.
Therefore, only the loop oscillation during the preflare phase is
detected. Finally, we want to stress that the low-amplitude
transverse oscillation is only observed in the wavelength of
AIA~171~{\AA}, and it is most like a kink oscillation in the coronal
loop \citep[e.g.,][]{Su12,Nistico13,Anfinogentov15,Li18}, but we
could not rule out the other MHD modes here.

\begin{figure}
\epsscale{0.9} \plotone{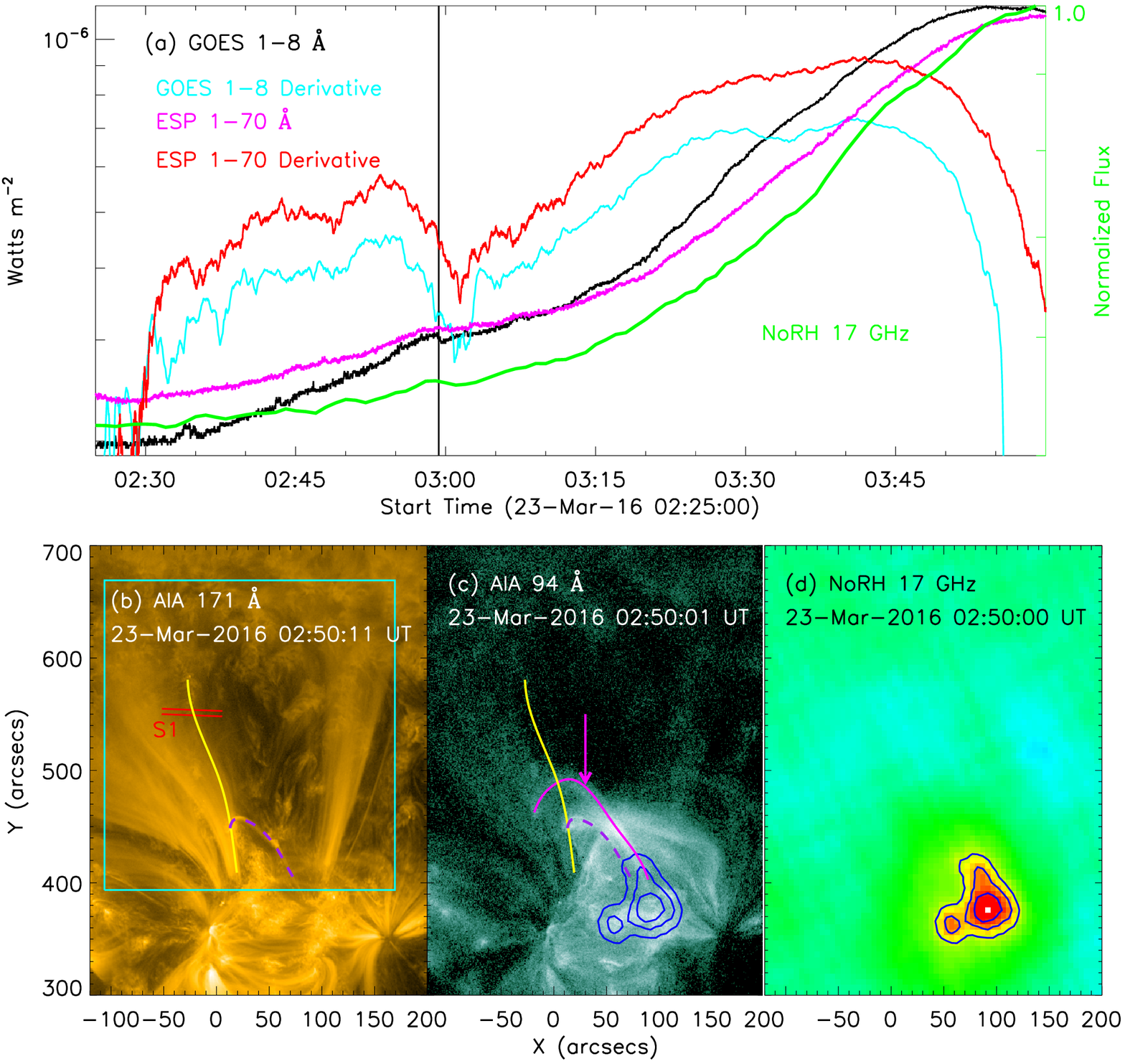} \caption{Upper: SXR light curves on
2016 March 23 in {\it GOES} 1$-$8~{\AA} (black), ESP~1$-$70~{\AA}
(magenta) and their time derivatives, as well as the normalized flux
at 17 GHz (green) integrated from the active region (panel~d)
measured by the NoRH. The black vertical line indicates the onset
time of the solar flare. Bottom: Nearly simultaneous images
($\sim$320\arcsec$\times$400\arcsec) recorded by {\it SDO}/AIA and
NoRH. The cyan box outlines the FOV of the movie (171\_94.mp4), and
two red lines mark the boundaries of slice (S1) used to plot the
time-distance images. The yellow line marks part of the coronal
loop, and the purple dashed line indicates a sheared arch, while the
magenta line (or arrow) indicates a hot loop in AIA~94~{\AA}. The
blue contours represent the microwave emissions at the levels of
70\%, 80\%, 90\%.} \label{flux}
\end{figure}

It is necessary to stress that a similar small-scale QPP is found in
microwave emission, i.e., the growing periods from $\sim$300~s to
$\sim$500~s during the preflare phase, and the QPP emission is very
weak. Here, the microwave light curve is integrated over the active
region (Figure~\ref{flux}~d), but not over the full solar disk.
Since the microwave radiation before a {\it GOES} C1.1 flare is very
weak, and the full-disk light curve show a strong background
emission. On the other hand, the time derivatives in SXR channels
recorded by {\it GOES}~1$-$8~{\AA} and ESP~1$-$70~{\AA} show a
significant enhancement during $\sim$02:30$-$03:01~UT, which closely
matches the microwave light curve. Our result is similar to previous
observations of the relationship between the SXR derivative light
curves and the microwave fluxes in solar flares
\citep[e.g.,][]{Neupert68,Ning08}. All these observational facts
suggest a thermal process during the preflare phase. Although the
microwave emission at NoRH 17~GHz might be the nonthermal
gyrosynchrotron radiation, it also could arise from the thermal
bremsstrahlung and gyroresonance \citep[e.g.,][]{Warmuth16}.
Therefore, the QPP in microwave channel may be triggered by the
thermal heating process, such as the kink mode.

\begin{figure}
\epsscale{1.0} \plotone{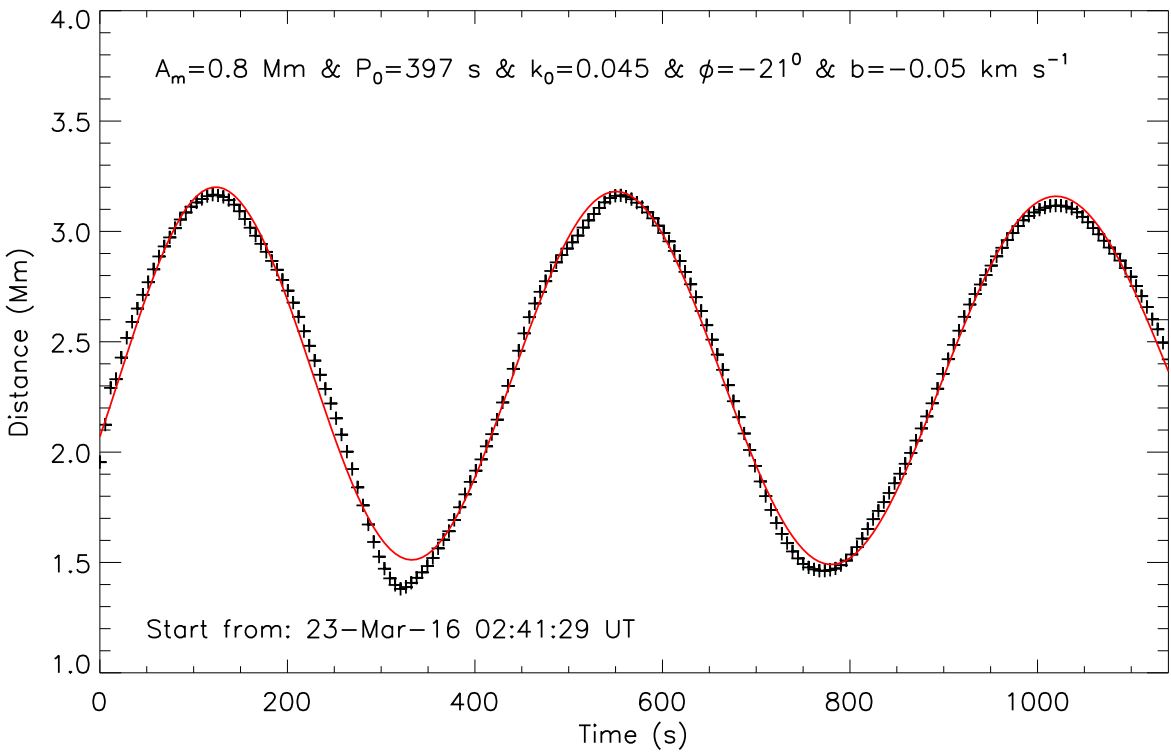} \caption{Oscillation locations
(pluses) of the diffuse coronal loop and their best fitted curve
(red).} \label{fitp}
\end{figure}

\begin{figure}
\epsscale{1.0} \plotone{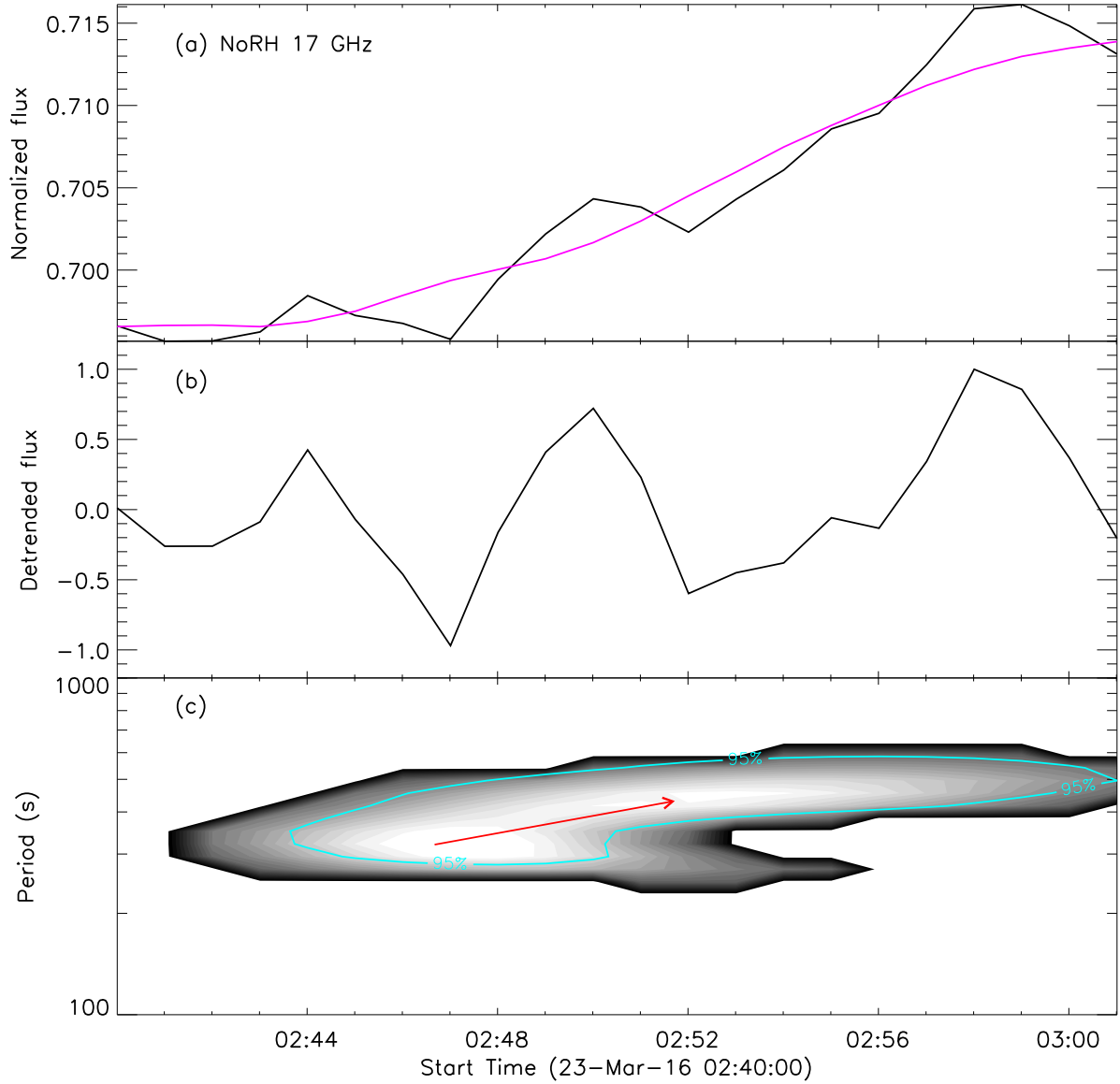} \caption{Panel~(a): Normalized
microwave light curve recorded by NoRH (black) and its trended flux
(magenta). Panel~(b) and (c): The detrended light curve and its
wavelet power spectrum. The cyan line represents a significance
level of 95\%, a red arrow indicates the growing period.}
\label{wavlet}
\end{figure}

There are three significant peaks in the low-amplitude transverse
oscillation of coronal loop in AIA~171~{\AA}, and the similar three
peaks can be discovered in the small-scale QPP of microwave emission
recorded by NoRH~17~GHz. Moreover, they exhibit a well one-to-one
corresponding relationship, as shown in Figure~\ref{slice}. Their
oscillation periods appear a clear growth trend, as presented in
Figures~\ref{fitp} and \ref{wavlet}. Thus, there must be some
connections between the loop oscillation and microwave QPP. Then, a
hot loop is found to gradually appear in AIA image sequences at
wavelength of 94~{\AA}, and it connects the EUV diffuse coronal loop
and microwave radiation source, as can be seen in the movie of
171\_94.mp4. The hot loop is much lower than the diffuse coronal
loop in AIA~171~{\AA}. This is consistent with the hot channel in
AIA~131~{\AA}, which is also much lower than the cool compression
front in AIA~171~{\AA} \citep[e.g.,][]{Zhang12,Cheng14}. On the
other hand, a highly sheared arch is detected during the preflare
phase in AIA~171~{\AA}, which is lower than the hot loop in
AIA~94~{\AA}, as shown in Figures~\ref{flux}. The observations
suggest that a weak magnetic reconnection process occurs before the
C1.1 flare, resulting into a slowly rising diffuse loop in
AIA~171~{\AA} \citep[see,][]{Liy17}. However, it is not necessary of
nonthermal process to generate QPPs, particularly for the
low-amplitude transverse oscillation with a period of several
minutes \citep[e.g.,][]{Tan10}. On the other hand, the solar
emissions of SXR, EUV, and 17~GHz microwave radiation can be related
to thermal processes. Therefore, the small-scale QPP in the EUV
coronal loop and microwave flux could be related to the same thermal
heating process in the oscillating source region during the preflare
phase \citep{Nakariakov09,Tan10}. Their oscillation periods might be
modulated by a kink oscillation in the EUV coronal loop
\citep{Tian12,Nistico13,Anfinogentov15}.

Finally, we want to stress that both the low-amplitude transverse
oscillation in the EUV coronal loop and QPP at 17~GHz microwave
emission exhibit slowly growth rate of periods. One possible
mechanism for this QPP event with growing periods is the LRC-circuit
model, which is related to the longitudinal current in the coronal
loop \citep[e.g.,][]{Zaitsev98,Tan10,Tan16,Chen19}. In this model,
the decrease of the current may lead to the growing period of the
oscillation. The existence of the longitudinal current is a
signature of non-potentialities in the source region, and the QPP
with growing periods in the preflare phase may provide valuable
information for understanding the triggering process of solar
flares.

\acknowledgments  We acknowledge the anonymous referee for valuable
comments. The authors would like to thank Prof.~V.~M.~Nakariakov,
Drs. W.~Su, L.~P.~Li, and D.~Yuan for their inspiring discussions.
We appreciate the teams of {\it SDO}, {\it GOES}, and NoRH, for
their open data use policy. This study is supported by NSFC under
grants 11973092, 11873095, 11790300, 11790302, 11729301, 11773079,
the Youth Fund of Jiangsu No. BK20171108, as well as the Strategic
Priority Research Program on Space Science, CAS, Grant No.
XDA15052200 and XDA15320301. D.~Li is also supported by the
Specialized Research Fund for State Key Laboratories. The Laboratory
No. 2010DP173032.

\end{document}